\documentclass[doublecol]{epl2}
\usepackage{graphics}
\usepackage{graphicx}
\usepackage{dcolumn} 
\usepackage{bm}
\usepackage{epsfig}
\usepackage[latin1]{inputenc}
\newcommand{\bnoo}{Ba$_2$NaOsO$_6$}

\title{
      Orbital-quenching-induced magnetism  
  in  Ba$_2$NaOsO$_6$ }
\author{K.-W. Lee and W. E. Pickett} 
\institute{Department of Physics, University of California, Davis, 
      California 95616, USA}
\abstract{The double perovskite \bnoo~with heptavalent Os ($d^1$) is observed to
remain in the ideal cubic structure ({\it i.e.} without orbital ordering)
despite single occupation of the
$t_{2g}$ orbitals, even in the ferromagnetically ordered phase below
6.8 K.  
Analysis based on the {\it ab initio} dispersion expressed in terms of
an Os $t_{2g}$-based Wannier function picture,
spin-orbit coupling, Hund's coupling, and strong Coulomb repulsion
shows that the magnetic OsO$_6$ cluster is near a moment-less 
condition due to spin and orbital compensation.  Quenching
(hybridization) then drives the emergence of the small moment.
This compensation, unprecedented in transition metals, arises in a
unified picture that accounts for the observed Mott insulating behavior.}

\begin{document}
\maketitle

Orbital physics in transition metal oxides has attracted a good deal of
attention for three decades, with much of the activity focused on coupling
to spin, charge, and lattice degrees of freedom in $d^1$ systems.  The $d^1$
configuration is typically found in the early $3d$ transition metals (TMs),
{\it i.e.} titanates and vanadates, and the associated phenomena -- often
revolving around the tendency to orbitally order -- are 
remarkably rich.  The $d^1$ configuration can also occur 
in the mid- to late-$5d$
TM ions, which are distinguished also by unusually 
high formal valence states. 

Unless fully itinerant,
partially filled $d$ shells 
lead to non-spherical ions that often are accommodated by orbital
ordering (OO), that is, ordered alignment of the filled orbitals\cite{KK1,tokura}
in the manner
that minimizes strain, electronic, and magnetic energies.\cite{fang}  An outcome
is that OO has been identified as the driving mechanism in symmetry-breaking
structural and magnetic transitions,\cite{tokura} 
and may also be coupled to magnetism and charge
order, the Mott insulating $d^1$ perovskite 
YTiO$_3$ being a well studied example.\cite{eva} 
Very different behavior is shown by the $d^1$ system LaTiO$_3$ 
which has a structural distortion although
evidence of OO has been difficult to obtain.  A prominent explanation
is that orbital fluctuations dominate, leading to a disordered
{\it orbital liquid} ground state.\cite{oo1,oo2}

Double perovskite structure
Ba$_2$NaOsO$_6$ (BNOO) is a rare example of a heptavalent 
osmium compound, also rare because it is a ferromagnetic 
insulator\cite{stitzer,erickson,FM}
(T$_C$ = 6.8 K), and
it shows other perplexing behavior.
Although it has a single
$5d$ electron in the $t_{2g}$ complex that orders magnetically, 
it shows no evidence of orbital
order that would destroy its cubic symmetry.  On the other hand, the sister compound
La$_2$NaOsO$_6$ with high-spin $d^3$ Os configuration with nominally cubic
symmetry is observed to be highly distorted,\cite{gemmill} which is
ascribed to misfit arising from
the small cation radius.   The question of spin ordering surely is a
delicate one, since isostructural and isovalent Ba$_2$LiOsO$_6$ (BLOO)
orders {\it anti}ferromagnetically in spite of a very similar
Curie-Weiss susceptibility\cite{stitzer} and nearly identical volume.

The question of formal valence, and identification of several material
constants, can be identified from first principles local density approximation
(LDA) calculations
using two all-electron full-potential electronic methods
of FPLO and Wien2k.\cite{fplo1, fplo2, wien2k, basis}
The Fermi level E$_F$ lies within the $t_{2g}$ bands, confirming the heptavalent
nature of the Os ion.  A crystal (ligand) field splitting of almost 5 eV
separates the centroids of the $e_g$ and $t_{2g}$ bands, reflecting the
unusually strong $5d-2p$ hybridization, and a gap of roughly 
1.5 eV separates the
$t_{2g}$ symmetry bands from the more tightly bound
O $2p$-dominated OsO$_6$ cluster orbitals, with
gaps separating various degenerate molecular orbitals.\cite{djs1,djs2}

\begin{figure}[tbp]
\rotatebox{-90}{\resizebox{5.5cm}{7.5cm}{\includegraphics{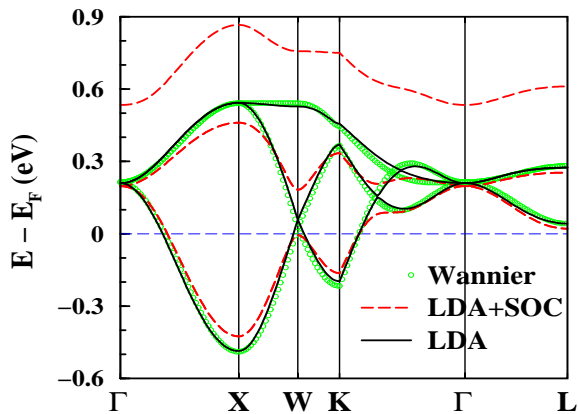}}}
\caption{(color online) Nonmagnetic Os $t_{2g}$ band structure
 in LDA, in three Wannier orbital representation (tight binding [TB]), 
and LDA+SOC, relative
to the Fermi level E$_F$.
 In LDA, the top of the $t_{2g}$ manifold is very flat near
 the $\Gamma$ point and along the $X$-$W$ line, resulting in
 very sharp peaks just above E$_F$ in
 the density of states (see fig. \ref{dos}).
 From LDA+SOC, the spin-orbit coupling constant $\xi$ is 0.30 eV.
}
\label{band}
\end{figure}

Several complications must be taken into account to obtain a 
realistic picture of the electronic structure.
(i) The strong hybridization of Os $5d$
orbitals with O $2p$ states, which has ramifications in this double
perovskite structure beyond the usual
ones in oxides.
(ii) Strong spin-orbit coupling (SOC) on the
Os site has important consequences. 
(iii) Hund's (exchange) coupling effectively selects the occupied orbital. 
(iv) Intra-cluster Coulomb repulsion, which is the obvious suspect
for a magnetic insulating state in an open shell system.
Finally, it is likely that quantum
fluctuations may also play an important role in determining the observed
behavior.  While the interplay of spin and orbital physics has been
addressed in considerable detail,\cite{oles} the large SOC in 
BNOO puts this compound
in a distinct class relative to the early $3d$ systems. 

\begin{figure}[tbp]
\rotatebox{-90}{\resizebox{6.5cm}{7.5cm}{\includegraphics{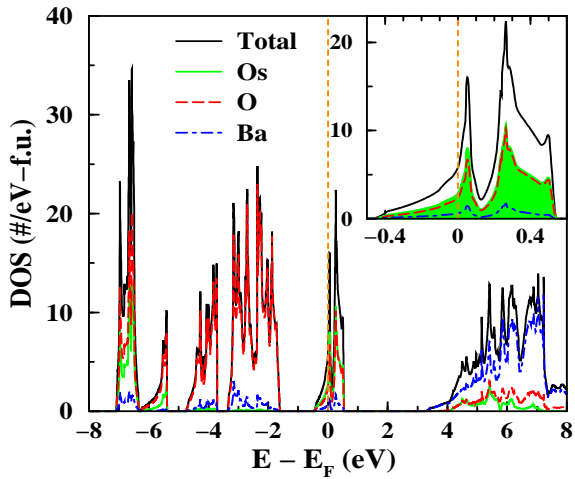}}}
\caption{(color online) Total and atom-projected densities of states
 (DOS) per formula unit for nonmagnetic state in LDA.
 The vertical dashed line indicates the Fermi energy.
 Inset: Blowup near E$_F$.
 It displays that contribution of six oxygens is nearly similar
 with Os character in the $t_{2g}$ manifold.
 As a result, it is anticipated that the oxygens significantly
 contribute to magnetic moment.
 }
\label{dos}
\end{figure}

\begin{figure}[tbp]
\rotatebox{-90}{\resizebox{6.5cm}{7.5cm}{\includegraphics{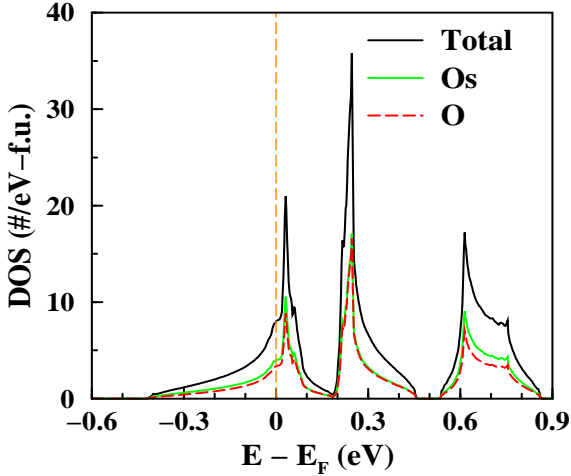}}}
\caption{(color online) Total and atom-projected DOS per formula unit,
with spin-orbit coupling included.  
Oxygen contribution is 50\% throughout the bands.
The two lower bands are distinct,
and nearly disjoint, leading roughly to a half-filled single-band
configuration. 
}
\label{pdos}
\end{figure}

The unusual aspects of BNOO then are: a Mott insulating state based on
a localized $d^1$ ion that nonetheless retains a cubic environment,
and a small ordered moment in a 3D system where fluctuation effects
typically are minor.
It is most instructive to consider the various aspects of the 
electronic structure consecutively.
We build an understanding of the electronic structure and competing
mechanisms in a different manner from Xiang and Whangbo\cite{whangbo}
in their study of BNOO.

\vskip 2mm  \noindent
{\it Isolated $t_{2g}$ bands: cluster orbitals.}
The $t_{2g}$ bands, with bandwidth $W$=1.05 eV,
form a self-contained system for the purpose
of understanding the low energy, low temperature physics.  The three
bands can be represented accurately in a Wannier function basis with three
first neighbor $d$-$d$ hopping parameters (in meV)
$t_{\sigma}=-121, t_{\pi}=64,
t_{\delta}=24$ meV, with on-site energy $\varepsilon=202$.  (For
BLOO $t_{\sigma}=-131$ is the only change.)
Note
that  $t_{\delta}$ is necessary to give the excellent representation of
the bands shown in
fig. \ref{band}.  The corresponding total and atom-decomposed densities of 
states (DOS) are pictured in fig. 2, which reveals the isolation of the 
narrow $t_{2g}$ bands.
As pointed out by
Erickson {\it et al.}\cite{erickson} the overlap takes place between 
O 2$p_{\pi}$ orbitals on neighboring OsO$_6$ octahedra; these O orbitals 
comprise the `tails' of the $t_{2g}$ symmetry cluster orbitals. The Fermi level 
lies 0.5 eV (one electron) up into the bands.
The DOS shown in fig. \ref{dos} is highly non-symmetric about their center, 
with 2/3 of the weight lying in the
upper half of the bands.

As analyzed in some detail by Singh and
coworkers\cite{djs1,djs2} in the context of pentavalent Ru
in Sr$_2$YRuO$_6$ which
also contains a closed shell ion (Y$^{3+}$) in one of the cation sites,
the OsO$_6$ unit should be approached as an isolated cluster.  The
Os $t_{2g}$ states generalize to cluster orbitals with the Os $5d_{xy}$
state (say) bonded to the $p_{\pi}$ orbitals lying in the $x-y$ plane.
We calculate that half of the density of the $t_{2g}$ bands lies on Os,
the other half distributed amongst four $p_{\pi}$ orbitals, and the cluster
orbital (with symmetric partners ${\cal D}_{xz}, {\cal D}_{yz}$) is
\begin{eqnarray}
{\cal D}_{xy} \approx \sqrt{\frac{1}{2}}d_{xy} +\sqrt{\frac{1}{8}}
   \sum_{j=1}^4 p_{\pi,j}.
\end{eqnarray}
This cluster orbital aspect, adopted also by Erickson and 
collaborators,\cite{erickson} is central to the following analysis.
The overlap of these cluster orbitals is what is described by
$t_{\sigma}, t_{\pi}, t_{\delta}$.

\vskip 2mm  \noindent
{\it Tendency toward magnetism.} 
Xiang and Whangbo reported\cite{whangbo} 
obtaining a FM ground state for BNOO.
Checking out this question with progressively finer 
$k$-point samplings of the zone,
we find BNOO is not quite unstable to a FM state.  For meshes of
8$\times$8$\times$8 or finer, the magnetic moment that is obtained
on coarser meshes vanishes.
Thus the Stoner interaction constant can be obtained from the condition
of (near) fulfillment of the Stoner instability:
$I_{st} = 1/N(E_F)$ =0.35 eV
[The Fermi level density of states N(E$_F$)=2.88 states/eV per spin].  
As will also be the case for the SOC strength (below),
it is important that this parameter
is not evaluated from the Os atomic potential alone, because the relevant local
(molecular) orbital is only half Os $5d$, and $I_{st}$ is reduced
accordingly. The Hund's (exchange) splitting
for a fully polarized $d^1$ orbital is ${\cal J}_H \approx I_{st}$ = 0.35 eV,
which is smaller than the bandwidth but comparable to the 
SOC strength (see below).

\vskip 2mm  \noindent
{\it Spin-orbit coupling.}
It has been known for fifty years\cite{stevens,enough,lacroix} that
the $t_{2g}$ complex
can be mapped onto an angular momentum $\vec L \rightarrow -\vec{\cal L}$
with quantum number ${\cal L}$=1, so within these bands the SOC operator becomes
$-\xi \vec {\cal L}\cdot \vec S$ with $\xi >$ 0.  
The basis $\{{\cal D}_{xy},{\cal D}_{yz},{\cal D}_{zx}\}$ can be mapped
onto the ${\cal L}_z$ eigenstates 
to $|0\rangle \rightarrow {\cal D}_{xy}; ~|\pm 1\rangle \rightarrow
{\cal D}_{zx} \pm i{\cal D}_{yz}$, where the
integer denotes the angular momentum projection $m_{\cal L}$ onto $\hat z$.
The negative sign leads
to an inverted spectrum, with the $J=\frac{3}{2}$ quartet at -$\xi$/2 and the
$J=\frac{1}{2}$ doublet at +$\xi$. 
The calculated splitting at $k$=0
gives a large SOC strength $\xi$=0.30 eV, 
so large it splits off the entire
upper $J$=1/2 band as is clear from fig. \ref{band}.

The DOS with SOC included is displayed in fig. 3.  In addition to the 
split-off upper $J=\frac{1}{2}$ band, the lower two bands
are nearly disjoint,  being bound together only by the degeneracy
within the $J=\frac{3}{2}$ states at zero momentum(fig. 1).  
As a result, the problem
of the Mott insulating state does not rigorously reduce to a single band
problem though it may be effectively single band.  
As a result of the mapping $\vec L
\rightarrow -\vec{{\cal L}}$ within the $t_{2g}$ states of interest, the total
angular momentum is $\vec J = \vec S - \vec {\cal L}$ 
($J = \frac{1}{2}, \frac{3}{2}$), the
magnetic moment is $\vec M = \mu_B ( 2\vec S - \vec{\cal L})$, and 
the $g$-factor is $g_J = \frac{3}{2} - \frac{5}{8J(J+1)}$.
In this isolated cluster limit, the r.m.s. moment $(\vec M^2)^{1/2}$
is small (1 $\mu_B$) for $J=\frac{1}{2}$ but is larger 
(${\sqrt 7} \mu_B$) in the
$J=\frac{3}{2}$ ground state.
  
At the band structure level, there remains only the Hund's
exchange splitting $H_{ex} = - {\cal J}_H S_z$ due to
the Os moment, which becomes important because
this interaction lifts the remaining degeneracies.  
The lowest state at $k$=0
is 
\begin{eqnarray}
|m_{\cal L} m_S\rangle = |+1\uparrow\rangle 
                   \propto |[{\cal D}_{zx} +i{\cal D}_{yz}]
                           \uparrow\rangle.
\end{eqnarray}
but the dispersion leads to overlapping bands and
metallicity, so intra-cluster Coulomb repulsion 
effects are required to describe the
observed insulating character, as also concluded by Xiang and Whangbo.

\vskip 2mm  \noindent
{\it Intra-atomic versus intra-molecular repulsion: the Mott insulating state.}
For a lattice of $d^1$ ions in $t_{2g}$ orbitals,  
an on-site Hubbard repulsion of roughly $U \sim W$ is expected to lead to a Mott
insulating state.  To model this insulating state we initially applied the
correlated band LDA+U method, using $J$=1 eV (whose value seemed
immaterial) and varying $U$.
Since the initial interest was to look for an orbitally ordered
solution, the full rotationally-invariant form of interaction 
implemented in Wien2k was
used, and (following conventional practice) the $U$ (and $J$) interaction is
applied only to the Os $5d$ orbitals.  A Mott insulating state was 
{\it not obtained} for values of $U$ even several times larger than $W$.
At $U$=4 eV $\approx$ 4$W$ the increasing exchange splitting leads 
only to a half metallic state, not an insulating state.  

This failure of the LDA+U method, as conventionally applied to atomic
orbitals, to produce the expected Mott insulating
state is a situation we
have observed previously in $4d$ and $5d$ oxides.  The problem can be cast 
not as a failure of the LDA+U
approach $per~se$, but rather of its application to an extended
{\it cluster} orbital by applying the orbital potentials solely to
the Os $5d$ {\it atomic} function, rather
than the entire molecular orbital ${\cal D}$.  Since only
50\% of the ${\cal D}$ orbital's charge is Os $5d$, the LDA+U correction
tends mainly to change the relative amounts of $5d$ and $2p$ character,
rather than to open a gap by shifting the energy of a cluster orbital.
As a result, no
state with integer $5d$ occupation number can arise, whereas
in the LDA+U method integer occupation is
necessary to obtain the insulating state.  This problem is cured by 
using the Wannier basis, rather than the atomic state basis, in the
LDA+U method.

\begin{figure}[tb]
\rotatebox{-90}{\resizebox{5.5cm}{7.5cm}{\includegraphics{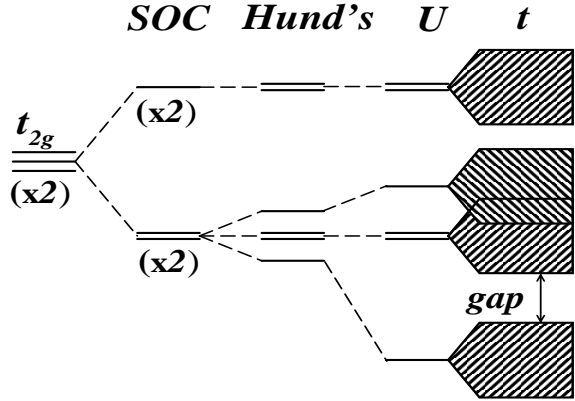}}}
\caption{ Evolution (schematic)
of the $t_{2g}$ energies from isolated
$3\times 2$-fold degenerate for the isolated ion (in its cubic environment),
through SOC, Hund's coupling, Coulomb repulsion $U$, and finally
dispersion due to hopping $t$.
The occupied band has pure spin-orbital $|+1\uparrow\rangle$ character
until intercluster hopping is included.
}
\label{tbband}
\end{figure}

\vskip 2mm  \noindent
{\it The full correlated electronic structure.}  The unfolding of the
electronic structure that we find most instructive is pictured
in fig. \ref{tbband}.
We begin from the basis of degenerate orthonormal cluster
orbitals ($H_0$), include SOC 
interaction $H_{SOC}$, then account for Hund's (exchange) interaction
${\cal J}$ that breaks the degeneracy, and finally the repulsive Coulomb 
interaction $H_{U}$ (as in LDA+U)
\begin{eqnarray}
 H
  &=& H_{0} + \xi\vec{L}\cdot\vec{S} - {\cal J}_H\hat{S_z} + H_{U}.
\label{eqn}
\end{eqnarray}
All of these are single site, and therefore simple to visualize.  
Kinetic effects (banding) is a small effect that enables charge transfer,
but when it does not destroy
the gap, as here, it is the least significant.

As mentioned above, Hund's exchange 
lowers the pure spin up $|+1\uparrow\rangle$ 
member of the lower spin-orbit
quartet, making it the orbital to occupy to achieve the lowest energy.
Then Coulomb repulsion lowers this occupied state by $\sim U$ with 
respect to the others.  
We find that a critical (minimum) value of $U_c = 1.1$ eV is required 
to produce the Mott insulating
state.  This critical value is significantly less than (therefore
consistent with) values that have been
suggested\cite{erickson,whangbo} as applicable to BNOO. This agreement reaffirms
that Coulomb repulsion on top of strong SOC is an essential aspect of the mechanism 
underlying the insulating state. The compensation of the moments 
distinguishes between the various viewpoints,
as we now describe.

The $|+1\uparrow\rangle$ state is pure-spin and pure-orbital moment,
and within this state the $z$-component
of the 
moment 
$\langle M_z\rangle 
 = \langle +1\uparrow|(2S_z - {\cal L}_z)|+1\uparrow\rangle \equiv 0$ 
is compensated precisely by opposing spin and orbital contributions. 
This feature makes BNOO an unprecedented
transition metal analog
of the case of the
Sm$^{3+}$ $d^5$ ion\cite{nature} where $L=2S$ and $M_z \sim 0$: 
the ion/cluster is magnetic ($J\neq 0$) but there is nearly vanishing 
total moment for a fixed direction of spin.  The effect of
mixing-in of other states by the kinetic hopping (quenching) is measured
by the resulting exchange coupling $4\frac{t^2}{U}\sim 30$ meV ($t\approx$
120 meV, $U$=1.5-2 eV), {\it i.e.} it is small compared to other energies
in the system.   While the 
spin and orbital moments each depend fairly strongly on the value of 
$U$ (table \ref{table2}), the cancellation between the two remains near-complete.
This cancellation is a nontrivial effect, since appreciable mixing-in
of the $|0\downarrow\rangle$ state along (say)
would degrade the spin moment without
affecting the orbital moment.

\begin{table}[bt]
\caption{Values of spin, orbital, and total moments versus the value
of $U$, within the correlated $t_{2g}$ shell of BNOO. Increasing the
hopping $t$ parameters (by 50-100\%) reduces the individual moments
but hardly changes the net moment.
}
\begin{center}
\begin{tabular}{cccc}\hline\hline
~$U$~  &  ~~~$M_S$~~~ &~~~ $M_L$~~~ &~~~$M_{tot}$~~~
  \\\hline
1.1 &   0.714    & -0.717 & -0.003 \\
1.5 &   0.909    & -0.866 & +0.043 \\
2.0 &   0.976    & -0.935 & +0.041 \\
\hline\hline
\end{tabular}
\end{center}
\label{table2}
\end{table}

\vskip 2mm  \noindent
{\it Discussion.} 
The small ordered moment was one of the principle questions about BNOO.  From
the point of view presented here, a small moment is {\it expected}
and the question becomes rather: why is it as large as observed?
Naively, this near-pure spin state is unexpected, given the
strong SOC which is largely regarded as mixing ({\it not} separating out)
spin and orbital moments, 
nevertheless it is an occurrence that has recently 
been anticipated.\cite{eschrig}
At the level of description of the Hamiltonian eq. \ref{eqn} in mean
field approximation, we have a Curie-Weiss moment $\langle M^2
\rangle^{1/2}$ = $\sqrt{7} \mu_B$ and an ordered FM moment of $\approx 0.04
\mu_B$ (see table \ref{tbband}), which is only nonzero due to quenching.   
The agreement with the observed values (near 0.7 and 0.2
$\mu_B$, respectively) in absolute terms is not good, 
however one would expect a small ordered
moment of 0.2 $\mu_B$ to be difficult to reproduce in a first principles
manner.  The degree of collinearity of $\vec S$ and $\vec{\cal L}$, which
finally determines the moment, is sensitive
to the environment.  The magnitude of collinearity, 
$\parallel \vec{\cal L}\cdot
\vec S \parallel / (\parallel S\parallel \parallel {\cal L}\parallel)$, 
is only 40\% in the $J=\frac{3}{2}$ manifold 
anyway.  Adding a crystalline anisotropy term $\gamma {\cal L}_z^2$ 
decreases it rapidly, by a further 35\% for $\gamma=\xi$ and by 60\%
for $\gamma=2\xi$ (this is orbital moment quenching).  The quantum 
fluctuations are thus quite sensitive to the environment.

The matter of quantitative
agreement of moments is actually a broader question: the related compound
Ba$_2$CaOsO$_6$ (BCOO) with one more
electron and $S$=1 and a cubic crystal structure is more conventional, 
yet it has a Curie-Weiss 
moment\cite{yamamura} of 1.61 $\mu_B$ that is also far below the 
spin-only value 
of 2.83 $\mu_B$. The Re-based $d^1$ double perovskites Sr$_2$XReO$_6$,
X = Mg and Ca, on the other hand, experience much smaller SOC and 
have moments equal to the spin only values and display structural  
distortions as the conventional picture would suggest.  This distinction 
supports the involvement of strong spin-orbit coupling in
both the small moment and the ``restoration'' of cubic symmetry in 
the heptavalent osmates BNOO
and BLOO.

The magnetic ordering issue 
(BNOO is FM; isostructural and isovalent BLOO 
with almost identical volume is AFM) is a challenging and delicate question
involving several complications.  First, the {\it fcc} lattice 
itself can be considered one of the most frustrated.  Frustration on the
triangular lattice, with its edge-sharing triangles, is well known.
Symmetric tetrahedra are frustrated, and the pyrochlore lattice of 
vertex-sharing tetrahedra is a well studied case of strong frustration.
The $fcc$ lattice can be pictured as one of space-filling, face-sharing
tetrahedra. 

Second, spin coupling between the OsO$_6$ clusters in BNOO
proceeds through three distinct hopping processes
$t_{\sigma}, t_{\pi}, t_{\delta}$ between anisotropic 
${\cal D}_{xz}+i{\cal D}_{yz}$ cluster
orbitals, and the behavior of the Curie-Weiss susceptibility reflects a
net antiferromagnetic coupling.  Such coupling is strongly frustrated on
an fcc lattice, and the actual pattern of ordering can be a delicate problem.
The difference between FM BNOO and AFM BLOO,
based on some small detail, reflects the geometrical frustration and apparent
competing couplings that must be expected in this structure.  
Note that
the cousin S=$\frac{1}{2}$ compound Sr$_2$CaReO$_6$, even with the 
coherent structural distortion that relieves some of the frustration,
has been shown to be a spin glass system.\cite{SCRO}

Thirdly, orbital ordering is also
subject to frustration.\cite{khomskii} The surviving
cubic structural symmetry of the Os $d^1$ ion is seems remarkable,
but one must recognize that the ``ferromagnetism'' arising in a strongly
coupled spin+orbital moment system lowers the electronic symmetry of
the system.  The question of whether this degree of freedom remains
disordered (a combined `spin-orbital liquid' within the FM phase), 
and thus serves to restore effective cubic symmetry, is a
question that requires further study. In any case, frustration
and large spin-orbital coupling are key issues in Ba$_2$NaOsO$_6$.

This work was supported by DOE grant No. DE-FG03-01ER45876
and DOE's Computational Materials Science Network.


\begin{thebibliography}{99}
\bibitem{basis} In FPLO, basis orbitals were chosen such as Ba $(4d5s5p)6s6p5d$,
 Na $(2s2p)3s3p3d$, Os $(4d4f5s5p)6s6p5d$, and O $2s2p3d$.
 (The orbitals in parentheses indicates semicore orbitals.)
 In Wien2k, the basis size was determined by
 $R_{mt}K_{max}$=7.0 and APW sphere radii (2.5 for Ba, 2.14 for Na,
 1.86 for Os, 1.65 for O).
\end{thebibliography}
\end{document}